\documentclass[journal]{IEEEtran}
\ifCLASSINFOpdf
\else
\fi

\usepackage{algorithm}  
\usepackage{algpseudocode}  
\usepackage{amsmath} 
\usepackage{amssymb}
\usepackage{array}
\usepackage{booktabs}
\usepackage{cite}
\usepackage{color,soul}
\usepackage{enumitem}
\usepackage[export]{adjustbox}
\usepackage{fancyhdr}
\usepackage{float}
\usepackage{graphicx}
\usepackage{hyperref}
\usepackage{lineno}
\usepackage{listings}
\usepackage{mdwmath}
\usepackage{mdwtab}
\usepackage{multirow}
\usepackage{multicol}
\usepackage{rotating}
\usepackage{setspace}
\usepackage{soul}
\usepackage{url}
\usepackage{verbatim}
\usepackage{xspace}
\usepackage{subfigure}
\usepackage{caption2}

\usepackage{hyperref}
\usepackage{url}

\soulregister\cite7
\soulregister\ref7

\usepackage[textsize=scriptsize,backgroundcolor=yellow!40]{todonotes}

\begin{document}
%
\title{Dynamic Fusion-based Federated Learning for COVID-19 Detection}



\markboth{Journal of \LaTeX\ Class Files,~Vol.~14, No.~8, August~2015}%
{Shell \MakeLowercase{\textit{et al.}}: Bare Demo of IEEEtran.cls for IEEE Transactions on Magnetics Journals}
%
\author{
\IEEEauthorblockN{
Weishan Zhang\IEEEauthorrefmark{1},
Tao Zhou\IEEEauthorrefmark{1},
Qinghua Lu\IEEEauthorrefmark{2},
Xiao Wang\IEEEauthorrefmark{3},
Chunsheng Zhu\IEEEauthorrefmark{4}\thanks{Weishan Zhang, Qinghua Lu, and Chunsheng Zhu are the corresponding authors. Email: zhangws@upc.edu.cn, qinghua.lu@data61.csiro.au, chunsheng.tom.zhu@gmail.com},\\
Haoyun Sun\IEEEauthorrefmark{1},
Zhipeng Wang\IEEEauthorrefmark{1},
Sin Kit Lo\IEEEauthorrefmark{2},
Fei-Yue Wang\IEEEauthorrefmark{3}\\
}
\IEEEauthorblockA{\IEEEauthorrefmark{1}College of Computer Science and Technology, China University of Petroleum (East China), China\\}
\IEEEauthorblockA{\IEEEauthorrefmark{2}Data61, CSIRO, Australia\\}
\IEEEauthorblockA{\IEEEauthorrefmark{3}State Key Laboratory of Management and Control for Complex Systems, Institute of Automation, Chinese Academy of Sciences, Beijing, China\\}
\IEEEauthorblockA{\IEEEauthorrefmark{4}SUSTech Institute of Future Networks, Southern University of Science and Technology, China}\\
}



\IEEEtitleabstractindextext{%
\begin{abstract}
Medical diagnostic image analysis (e.g., CT scan or X-Ray) using machine learning is 
an efficient and accurate way to detect COVID-19 infections. However, sharing diagnostic images across medical institutions is usually not allowed due to the concern of patients' privacy. This causes the issue of insufficient datasets for training the image classification model. Federated learning is an emerging privacy-preserving machine learning paradigm that produces an unbiased global model based on the received updates of local models trained by clients without exchanging clients' local data. Nevertheless, the default setting of federated learning introduces huge communication cost of transferring model updates and can hardly ensure model performance when data heterogeneity of clients heavily exists. 
To improve communication efficiency and model performance, in this paper, we propose a novel dynamic fusion-based federated learning approach for medical diagnostic image analysis to detect COVID-19 infections. First, we design an architecture for dynamic fusion-based federated learning systems to analyse medical diagnostic images. Further, we present a dynamic fusion method to dynamically decide the participating clients according to their local model performance and schedule the model fusion-based on participating clients' training time.
In addition, we summarise a category of medical diagnostic image datasets for COVID-19 detection, which can be used by the machine learning community for image analysis. The evaluation results show that the proposed approach is feasible and performs better than the default setting of federated learning in terms of model performance, communication efficiency and fault tolerance.
\end{abstract}

}

\maketitle

\IEEEdisplaynontitleabstractindextext

%
\IEEEpeerreviewmaketitle

\section{Introduction}
%
%
%
%

\IEEEPARstart{T}{he} COVID-19 pandemic has introduced an unprecedented global crisis. The rapidly increasing number of COVID-19 cases leads to a severe shortage of test kits and calls for a more efficient and accurate way to diagnose COVID-19 infections. To address the issue of the shortage of test kits for COVID-19 diagnosis, researchers have been working on machine learning technologies, especially deep learning, using medical diagnostic images (e.g., CT scan or X-Ray). The model performance is heavily dependent on the training dataset size and diversity. However, 
data hungriness is a critical challenge due to the concern for data privacy. To protect patients' privacy, sharing medical data across medical institutions is not allowed, which causes the issue of insufficient datasets for model training.

The concept of federated learning was introduced by Google in 2016 as a new machine learning paradigm that produces an unbiased model while preserving data privacy ~\cite{DBLP:journals/corr/McMahanMRA16,2020SLR}. In each round of training, clients (e.g., organisations, data centers, or mobile/IoT devices) are selected to train a model using local data and send the updates of local models to a central server for aggregation without transferring any local raw data. 

\begin{figure*}[!t]
	\centering
	\includegraphics[width=0.7\textwidth]{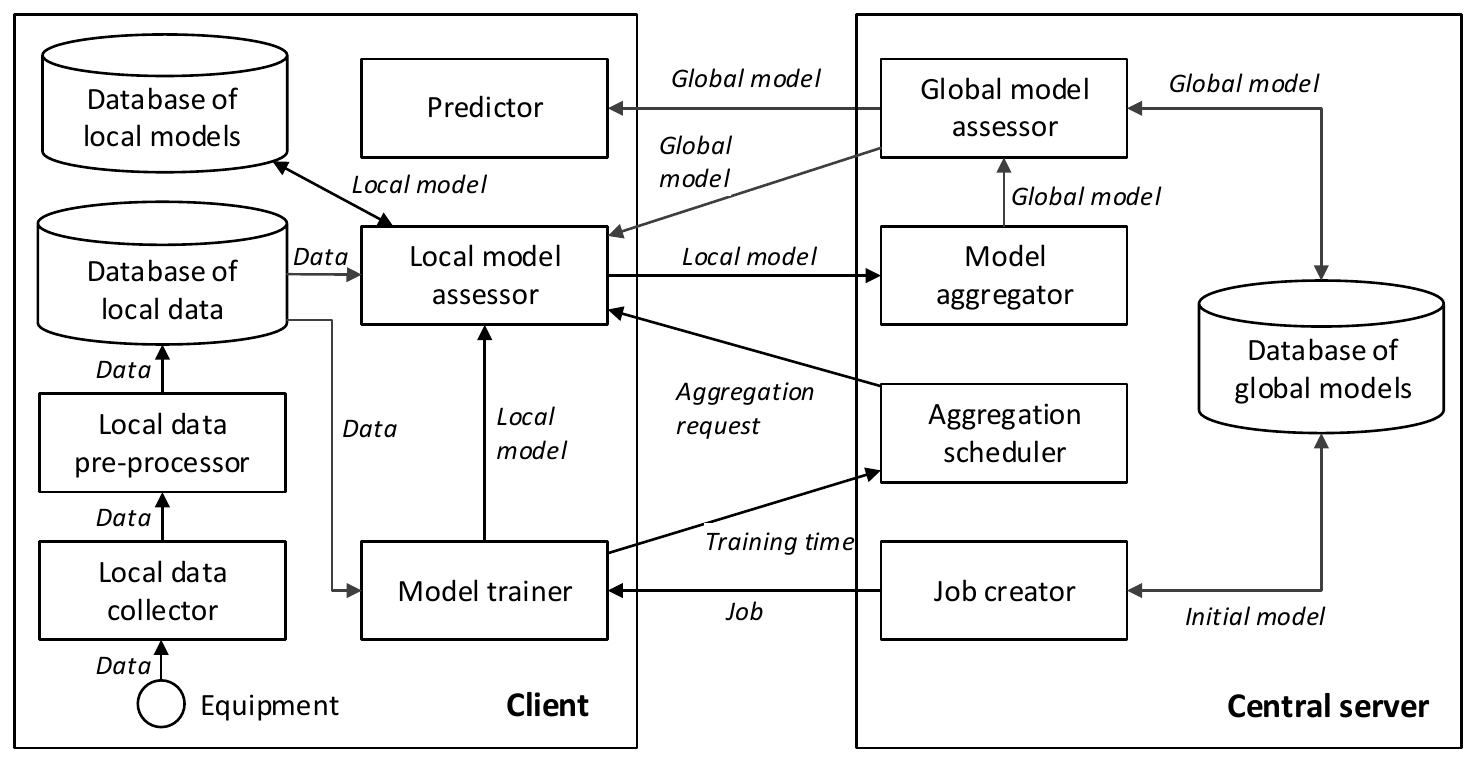}
	\caption{Architecture of federated learning systems for medical diagnostic image analysis}
	\label{softwarearchitecture}
\end{figure*}

Federated learning has the potential to connect isolated medical institutions and train a model for COVID-19 positive case detection while preserving data privacy. Some recent works leverage federated learning to diagnose COVID-19 infection through CT or X-Ray images~\cite{liu2020experiments, kumar2020blockchainfederatedlearning}. However, the above studies adopted the default setting of federated learning which might introduce huge communication cost of transferring model updates (e.g. massive matrices of weights) and under-performs when data heterogeneity of clients heavily exists. 

To improve communication efficiency and model performance, we propose a novel dynamic fusion-based federated learning approach for COVID-19 positive case detection. First, we design a dynamic fusion-based federated learning system architecture for medical diagnosis image analysis to detect COVID-19 positive cases. The proposed architecture provides a systematic view of the components' interactions and serves as a guide for the design of federated learning systems. Second, we present a dynamic fusion method to decide the participating clients according to their local model performance and schedule the model fusion dynamically, based on the participating clients' training time. Each client assesses the local model trained and only uploads the model updates when it performs better than the previous version 
while the central server configures the waiting time for each client to send model updates based on the average training time for the last round. Additionally, we summarise a category of medical diagnostic image datasets for COVID-19 detection, which can be used by the machine learning community for image analysis. The evaluation results show that the proposed approach achieves better detection accuracy, fault tolerance, and communication efficiency compared to the default setting of federated learning.


The remainder of this paper is organized as follow. Section \ref{architecture} presents the approach. Section \ref{evaluation} evaluates the approach. Section \ref{background} discusses the related work. Section \ref{conclusion} concludes the paper.

\section{Dynamic Fusion-based Federated Learning for COVID-19 Detection}
\label{architecture}

In this section, we present a dynamic fusion-based federated learning approach for CT scan image analysis to diagnose COVID-19 infections. Section \ref{arc} provides an overview of the architecture and discusses how the components and their interactions.  Section \ref{fusionmethod} discusses a dynamic model fusion method to dynamically decide the participating clients and schedule the aggregation based on each participating client's training time. 

\begin{figure*}[!t]
	\centering
	\includegraphics[width=0.6\textwidth]{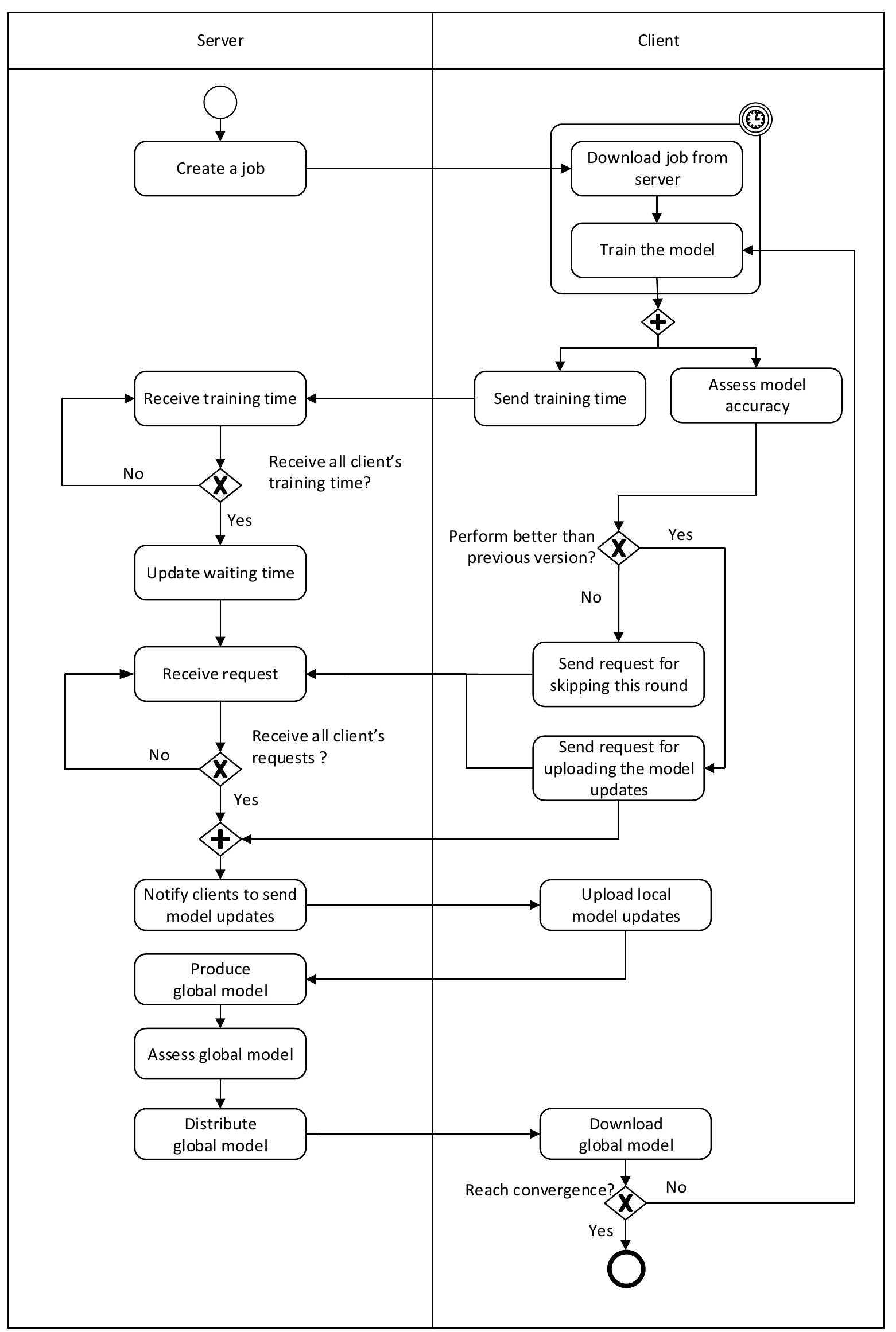}
	\caption{Dynamic fusion process}
	\label{fusion}
\end{figure*}

\subsection{Architecture}
\label{arc}
Fig.~\ref{softwarearchitecture} illustrates the architecture, which consists of two types of components: central server and clients. The central server initialises a machine learning job and coordinates the federated learning process, while clients train local models specified in the learning job using local data and computation resources.

Each client gathers images scanned by the diagnostic imaging equipment through the client data collector and cleans the data (e.g., noise reduction) via the client data pre-processor and store locally. 
The job creator initialises a model training job (including initial model code and the number of aggregation) and configures the initial waiting time for clients to return the model updates.
Each participating client downloads the job and 
trains the model via the model trainer. After a set number of epochs, the model trainer completes this round of training and uploads the training time to the central server. The aggregation scheduler updates the waiting time based on the training time received from participating clients. 

The local model assessor on each client compares the performance of the current local model with the previous version. If the current local model performs better, the client sends a request for model upload to the central server. Otherwise, the client will request to not upload the model update for this round. All the clients that do not complete the set number of epochs within the current waiting time are not allowed to participate in this aggregation round. After the set waiting time, the aggregation scheduler on the central server notifies the clients that have sent the model upload request. After the aggregation, the global model assessor measures the accuracy of the aggregated global model and sends the global model back to each client for a new round of training.


\begin{algorithm}[ht]
\caption{Dynamic fusion algorithm.}  
  \label{algorithm}
  \footnotesize
  \begin{algorithmic}[1]
  \State /*Client*/
  \State $Job \gets$ download($ServerURL$)
  \State $FusionTimes, Model \gets$ decode($Job$)
  \State $FedStep  \gets$ 0
  \While {$FedStep \textless$ $FusionTimes$}
  \State $Acc, TrainingTime \gets$ train($Model$)
  \State send($TrainingTime$)
  \State $MaxAcc \gets$ request($ServerURL$)
  \If {$Acc \ge$ $MaxAcc$}
  \State upload($Model$)
  \State $Model \gets$ receiveModel()
  \EndIf
  \State $FedStep ++$
  \EndWhile
  \\
  \State /*Server*/
  \State $WaitingTime, MaxAcc, FusionTimes, Model \gets$ initialize()
  \State $Job \gets$ encode($FusionTimes, Model$)
  \While {true}
  \State $TrainingTime \gets$ receive()
  \State $WaitingTime \gets$ update($TrainingTime$)
  \State $ClientModel \gets$ receiveModel()
  \If {expired($WaitingTime$) $== true$}
  \State $Model \gets$ aggregate($ClientModel$)
  \State $MaxAcc \gets$ evalate($Model$)
  \State dispatch($Model$)
  \EndIf
  \EndWhile
  \end{algorithmic}
\end{algorithm}

\subsection{Dynamic Fusion}
\label{fusionmethod}
To improve communication efficiency in federated learning, the proposed dynamic fusion method consists of two decision-making points: client participation and client selection.  
On the client side, each client decides whether to join this round of aggregation based on the performance of the newly trained model. 
On the central server side, the model aggregator selects the participating clients based on the waiting time
If a client does not upload the model update within the waiting time, it is excluded by the central server for this round of aggregation. The waiting time of current the round is calculated by averaging the previous round's training time of each client. The initial waiting time is configured by the platform owner.

Fig.~\ref{fusion} illustrates the process of the proposed dynamic fusion method, and Algorithm~\ref{algorithm} describes the detailed process. The process starts with creating a learning job by the central server. All the clients download the job from the central server and set up the local training environment. From the second round, a timer is set for each client based on the average training time of all the participating clients for the previous round. If a client does not complete the training within the configured time, the central server proceeds the aggregation without any input from this client for this round. On the other hand, if the model trained by the client for this round performs worse than last round, the client sends a request to the central server for skipping this round's aggregation. Otherwise, the client notifies the central server to update the model.

\section{Evaluation}
\label{evaluation}

\begin{table*}[tbhp]
\footnotesize
\setlength{\belowcaptionskip}{10pt}
\centering
\caption{A Category of Medical Diagnostic Image Datasets for COVID-19 Detection.}
\label{Dataset_1}
\begin{tabular}{p{0.15\columnwidth}p{0.13\columnwidth}p{0.13\columnwidth}p{0.13\columnwidth}p{0.13\columnwidth}p{0.1\columnwidth}p{0.93\columnwidth}}
\toprule

\multicolumn{1}{l}{\bf Type} &
\multicolumn{1}{l}{\bf Amount} &
\multicolumn{1}{l}{\bf Size} &
\multicolumn{1}{l}{\bf COVID-19} &
\multicolumn{1}{l}{\bf Negative} &
\multicolumn{1}{l}{\bf VP} &
\multicolumn{1}{c}{\bf Github Address}\\
\midrule

CT & 746 & 92.6M & 349 & 397 & 0 & \url{https://github.com/UCSD-AI4H/COVID-CT} \\
\cmidrule(l){1-7}

X-ray & 2905 & 1168M & 219 & 1341 & 1345 & \url{https://www.kaggle.com/tawsifurrahman/covid19-radiography-database} \\
\cmidrule(l){1-7}

X-ray & 55 & 14.2M & 55 & 0 & 0 & \url{https://github.com/agchung/Figure1-COVID-chestxray-dataset} \\


\bottomrule
\end{tabular}
\end{table*}

\begin{table}[tbhp]
\footnotesize
\setlength{\belowcaptionskip}{10pt}
\centering
\caption{Experiment environment.}
\label{ExperimentEnvironment}
\begin{tabular}{p{0.1\columnwidth}p{0.25\columnwidth}p{0.1\columnwidth}p{0.1\columnwidth}p{0.1\columnwidth}}
\toprule
\multicolumn{1}{l}{\bf Node}
 &
\multicolumn{1}{l}{\bf GPU} &
\multicolumn{1}{l}{\bf RAM} &
\multicolumn{1}{l}{\bf Python} &
\multicolumn{1}{l}{\bf CUDA} \\
\midrule

Server & RTX 2080Ti & 11G & 3.6 & 10.0 \\
\cmidrule(l){1-5}

Client1 & GTX 1070 & 8G & 3.6 & 10.1 \\
\cmidrule(l){1-5}

Client2 & GTX 1080 & 8G & 3.6 & 10.1 \\
\cmidrule(l){1-5}

Client3 & TITAN X(Pascal) & 12G & 3.7 & 10.1 \\

\bottomrule
\end{tabular}
\end{table}

\begin{table*}[tbhp]
\footnotesize
\setlength{\belowcaptionskip}{10pt}
\centering
\caption{Dataset Configuration for Each Client.}
\label{Dataset_2}
\begin{tabular}{p{0.15\columnwidth}p{0.15\columnwidth}p{0.15\columnwidth}p{0.15\columnwidth}p{0.15\columnwidth}p{0.15\columnwidth}p{0.15\columnwidth}p{0.22\columnwidth}p{0.15\columnwidth}}
\toprule

\multicolumn{1}{l}{\bf Client 1} &
\multicolumn{1}{l}{\bf Data Size (MB)} &
\multicolumn{1}{l}{\bf Client 2} &
\multicolumn{1}{l}{\bf Data Size (MB)} &
\multicolumn{1}{l}{\bf Client 3} &
\multicolumn{1}{l}{\bf Data Size} &
\multicolumn{1}{l}{\bf Ratio} &
\multicolumn{1}{l}{\bf Total Data Size (MB)} &
\multicolumn{1}{l}{\bf Amount}\\
\midrule

600/0 & 76.8 & 0/900 & 391.3 & 0/1300 & 545.7 & 600/2200 & 1013.8 & \multirow{9}{0.15\columnwidth}{2800}\\
\cmidrule(l){1-8}

300/300 & 168.5 & 0/900 & 392.5 & 0/1300 & 546.6 & 300/2500 & 1107.6 &\\
\cmidrule(l){1-8}

200/400 & 196.8 & 0/900 & 389.1 & 0/1300 & 534.5 & 200/2600 & 1120.4 & \\
\cmidrule(l){1-8}

150/450 & 209.9 & 0/900 & 381.6 & 0/1300 & 544 & 150/2650 & 1135.5 & \\
\cmidrule(l){1-8}

200/400 & 197.4 & 200/700 & 318.9 & 0/1300 & 557.5 & 400/2400 & 1073.8 & \\
\cmidrule(l){1-8}

200/400 & 198.6 & 200/700 & 317.2 & 200/1100 & 497 & 600/2200 & 1012.8 & \\

\bottomrule
\end{tabular}
\end{table*}
Table~\ref{Dataset_1} summarises a category of medical diagnostic image datasets for COVID-19 detection which include 746 CT images and 2960 X-ray images. 
The proposed approach is evaluated via quantitative experiments using the datasets as shown in Table~\ref{Dataset_1}. 
The 746 CT dataset includes 349 images of COVID-19 positive cases and 397 images of negative cases. 
The chest X-ray images are from two datasets. The first X-ray dataset has 2905 images which contain 219 images of COVID-19 positive cases, 1341 images of negative cases, and 1345 images of viral pneumonia (VP). The second X-Ray dataset consists of 55 images of positive cases.

As shown in Table~\ref{ExperimentEnvironment}, the experiments involve one central server and three clients with different configurations. We selected 3326 images from the collected datasets and divided them into 2800 images for the training set and 526 images for the test set.
We set different dataset sizes for each client: 600 images, 900 images, and 1300 images respectively. Considering the difference between CT and X-ray images, we adjusted the ratio of these two types of images while keeping the same total amount for each client, which is shown in Table~\ref{Dataset_2}. In the test set, there are 71 CT images (31 COVID-19 positive cases, 40 negative cases), and 455 X-ray images (55 COVID-19 positive cases, 200 negative cases, and 200 virus pneumonia). Please note that the CT images are taken from the top, while the X-Ray images are taken from the front.

\subsection{Accuracy}


To evaluate the accuracy of dynamic fusion-based federated learning (DF\_FL), we conducted experiments using three different models, GhostNet, ResNet50, and ResNet101. The models were trained with the six groups of datasets listed in Table~\ref{Dataset_2}. There are 18 groups of experiments in total. We compared the results with the default setting of federated learning (D\_FL). GFL federated learning framework\footnote{\url{https://github.com/GalaxyLearning/GFL}} was used in our experiments.

\begin{figure*}[!ht]
    \centering
	\includegraphics[width=1.4\columnwidth]{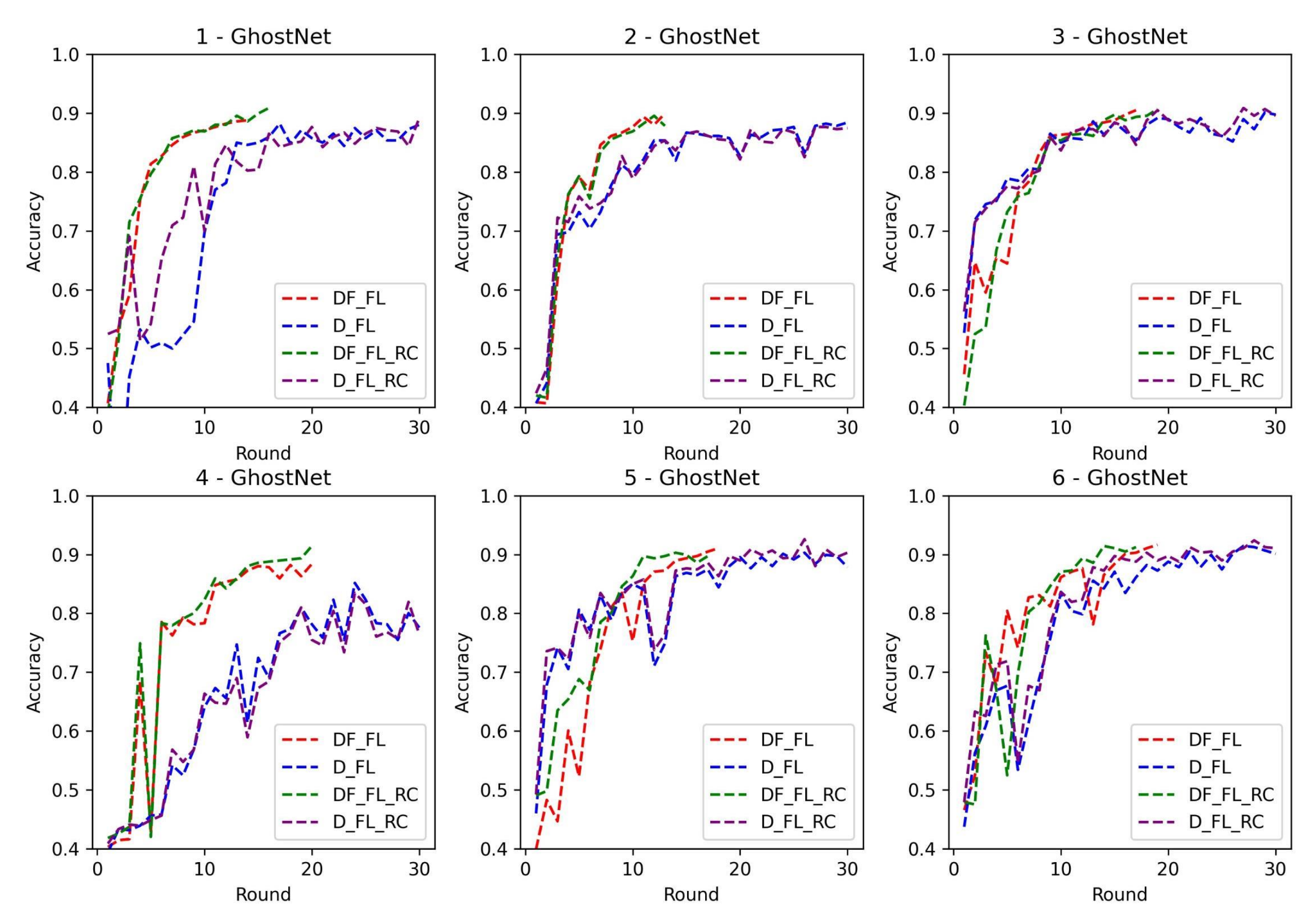}
	\caption{Accuracy of GhostNet.}
	\label{Accuracy_GhostNet}
\end{figure*}

\begin{figure*}[!ht]
    \centering
	\includegraphics[width=1.4\columnwidth]{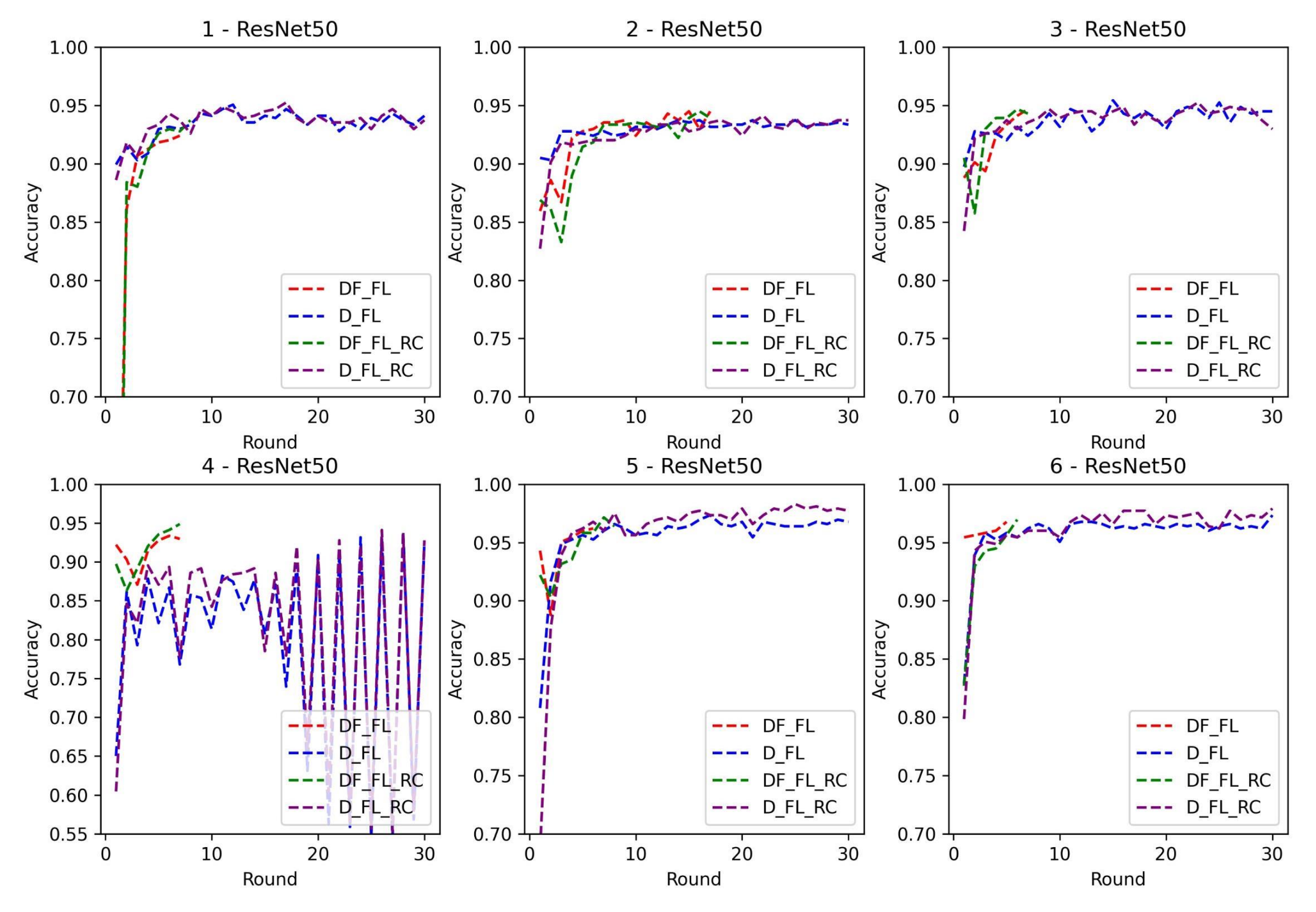}
	\caption{Accuracy of ResNet50.}
	\label{Accuracy_ResNet50}
\end{figure*}

\begin{figure*}[!ht]
    \centering
	\includegraphics[width=1.4\columnwidth]{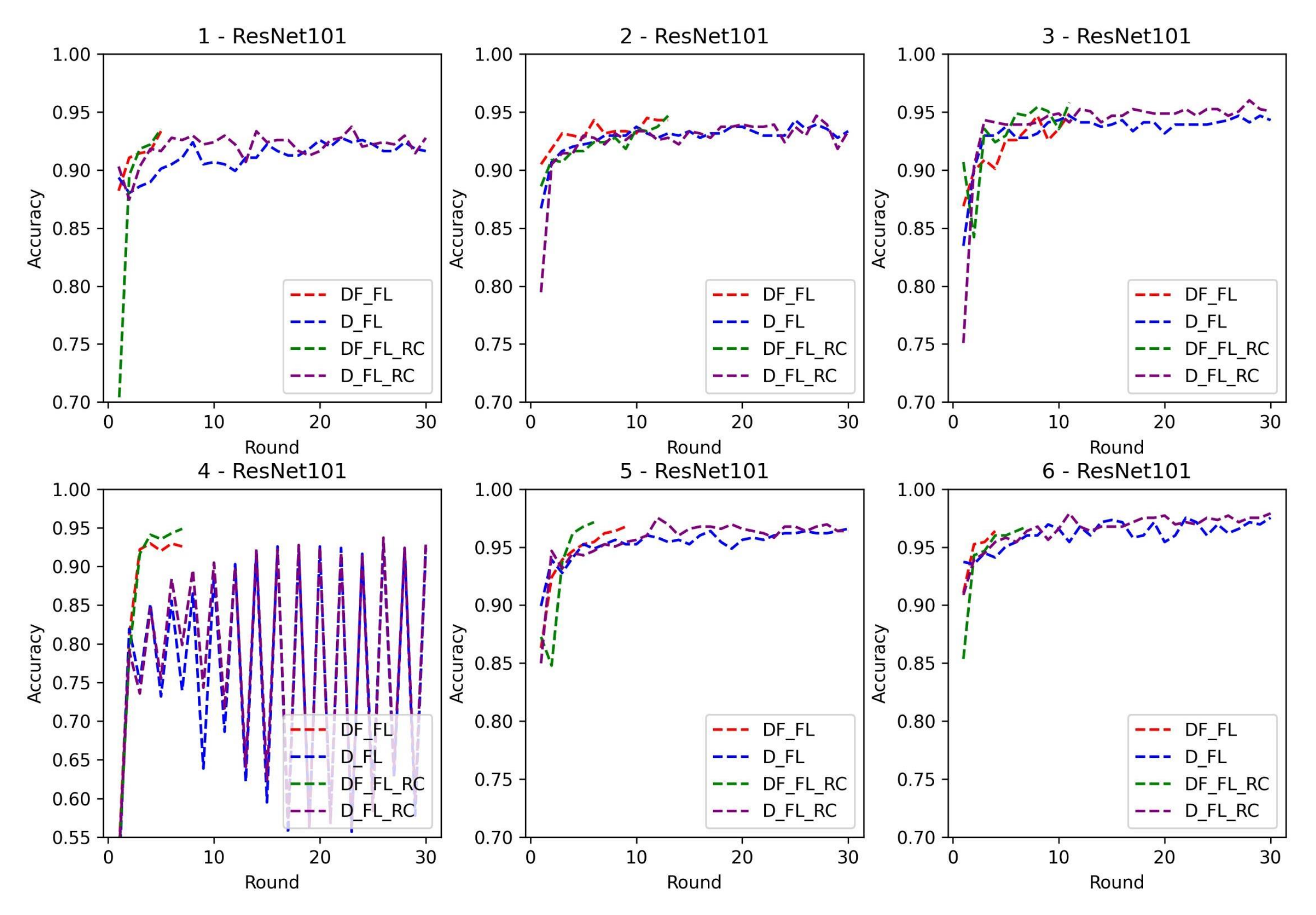}
	\caption{Accuracy of ResNet101.}
	\label{Accuracy_ResNet101}
\end{figure*}

The results are presented in Fig.~\ref{Accuracy_GhostNet}, Fig.~\ref{Accuracy_ResNet50}, Fig.~\ref{Accuracy_ResNet101} respectively for each type of model. The results show that in the 18 groups of experiments, there are only 4 groups in which the dynamic fusion-based federated learning (DF\_FL) achieves lower accuracy than the default setting of federated learning (D\_FL) (lower than D\_FL by 1.711\%, 0.57\%, 0.57\%, and 1.141\% respectively). 14 groups in which the dynamic fusion-based federated learning (DF\_FL) achieves higher accuracy than the default setting of federated learning (D\_FL). 
Overall, the proposed dynamic fusion-based federated learning approach achieved higher accuracy compared to the default setting of federated learning. Also, an interference is introduced in the 4th group of the dataset for each model, where images of negative cases are marked as positive COVID-19. The model trained by fusion-based federated learning can still achieve relatively steady results and higher accuracy compared to the default setting, which shows that the proposed fusion-based federated learning can ensure fault tolerance and robustness.






















In addition, we measured the accuracy of each type of model using the test set which was processed by random cropping. The results are also shown in Fig.~\ref{Accuracy_GhostNet}, Fig.~\ref{Accuracy_ResNet50}, and Fig.~\ref{Accuracy_ResNet101}. Similarly, the results demonstrate that the proposed dynamic fusion-based federated learning (DF\_FL) achieves higher accuracy than the default setting of federated learning (D\_FL) in 14 groups of experiments. For the rest, DF\_FL is lower than D\_FL for 0.57\%, 1.331\%, 0.951\%, and 1.141\% respectively
The results show that the proposed fusion-based federated learning perform better in real-world datasets than the default setting of federated learning.

\subsection{Training Time}
\begin{figure}[t]
	\centering
	\subfigure[GhostNet]{
	\centering
    \includegraphics[width=0.8\columnwidth]{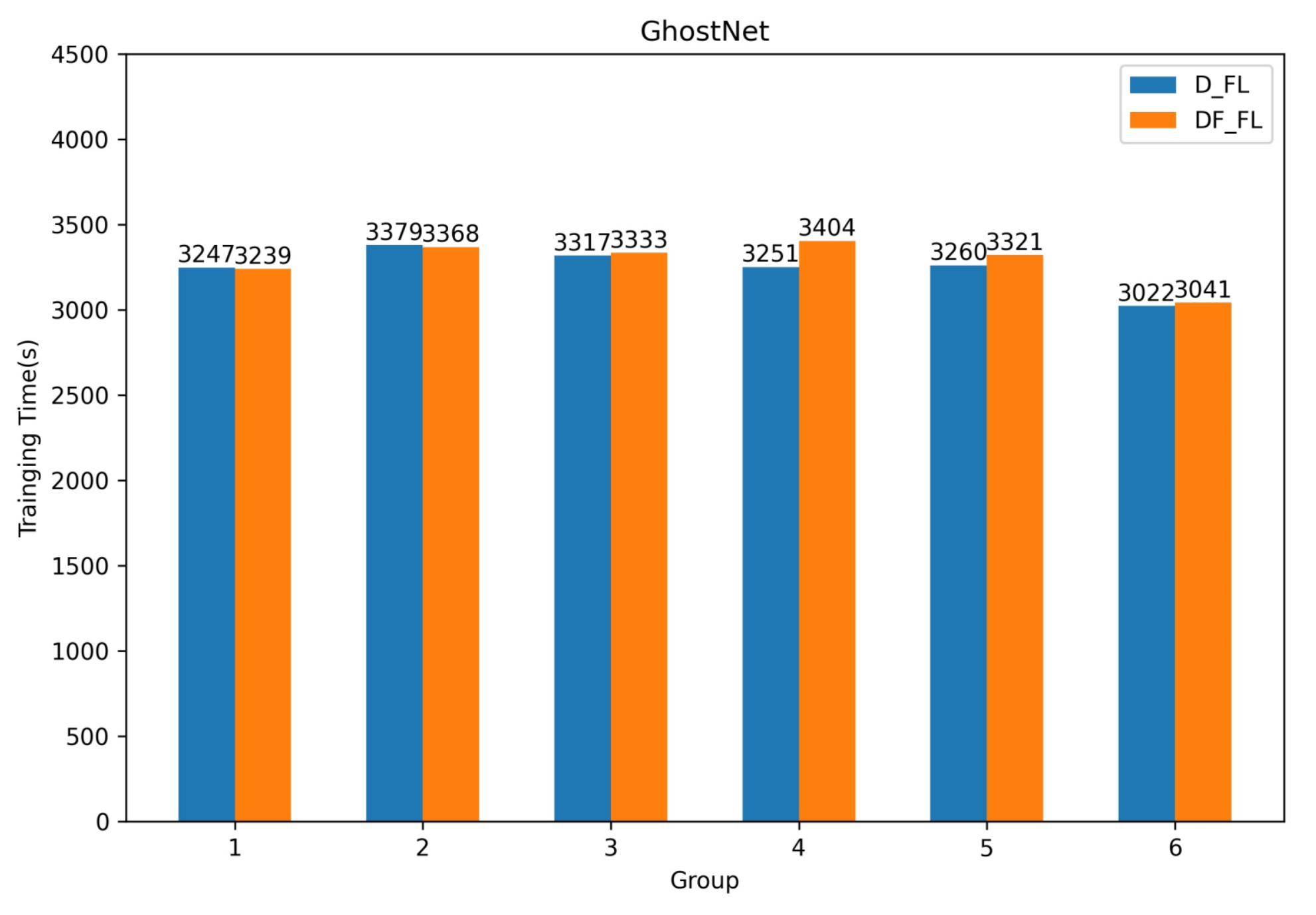}
    }
    \qquad
	\subfigure[ResNet50]{
	\centering
    \includegraphics[width=0.8\columnwidth]{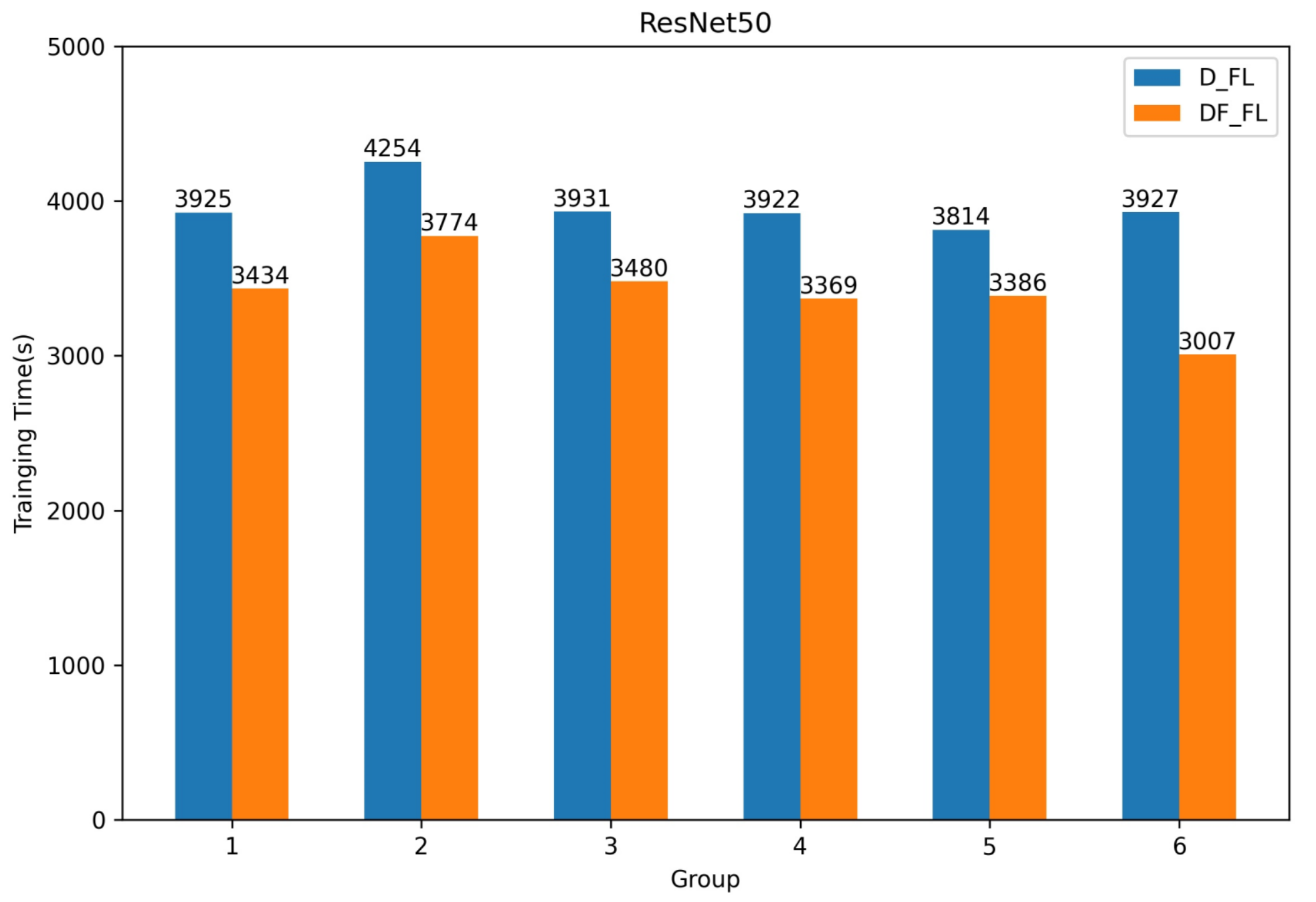}
    }
    \qquad
    \subfigure[ResNet101]{
    \centering
    \includegraphics[width=0.8\columnwidth]{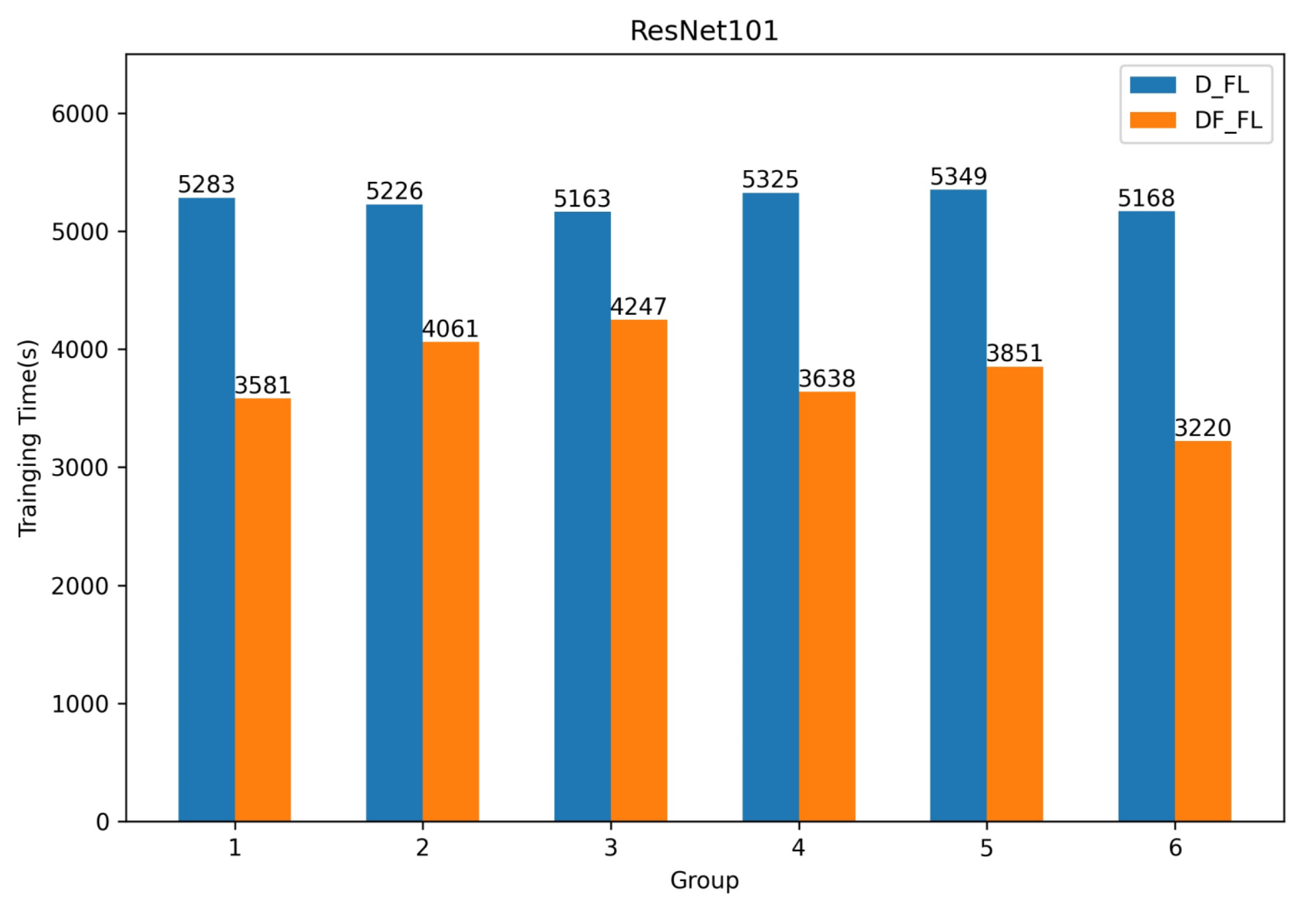}
    }
    \caption{Training time.}
	\label{training_time}
\end{figure}

To evaluate the training efficiency of the proposed dynamic fusion-based federated learning, we recorded the training time during the above experiments. The training epochs of the clients are set to 90 and the maximum network speed is configured as 10 MB/s for model upload/download (10MB/s). The recorded training time is illustrated in Fig.~\ref{training_time}. The results show that in GhostNet, the proposed dynamic fusion-based federated learning does not lower the training time, while there is an apparent effect on ResNet50 and ResNet101. The training time of ResNet50 is reduced by 8-10 minutes, while the training time of ResNet101 is decreased by 25-30 minutes. 

Since we found that the proposed dynamic fusion-based federated learning cannot reduce the training time of GhostNet network in the above experiments, we further study the influence factor. After measuring the single model transmission time, 
we observe that the GhostNet has less parameters compared to the other two networks. Thus, GhostNet costs less time for model transmission (which is 2.2s on average), which results in no change in GhostNet training time. In contrast, ResNet50 and ResNet101 have more parameters that take more time to transmit the model updates. Thus, there is an apparent improvement in these two networks in terms of communication efficiency. We can conclude that applying the proposed dynamic fusion-based approach can significantly reduce the training time when the network is poor and the model has large amounts of parameters.

\begin{figure}[t]
	\centering
	\subfigure[GhostNet]{
	\centering
     \includegraphics[width=\columnwidth]{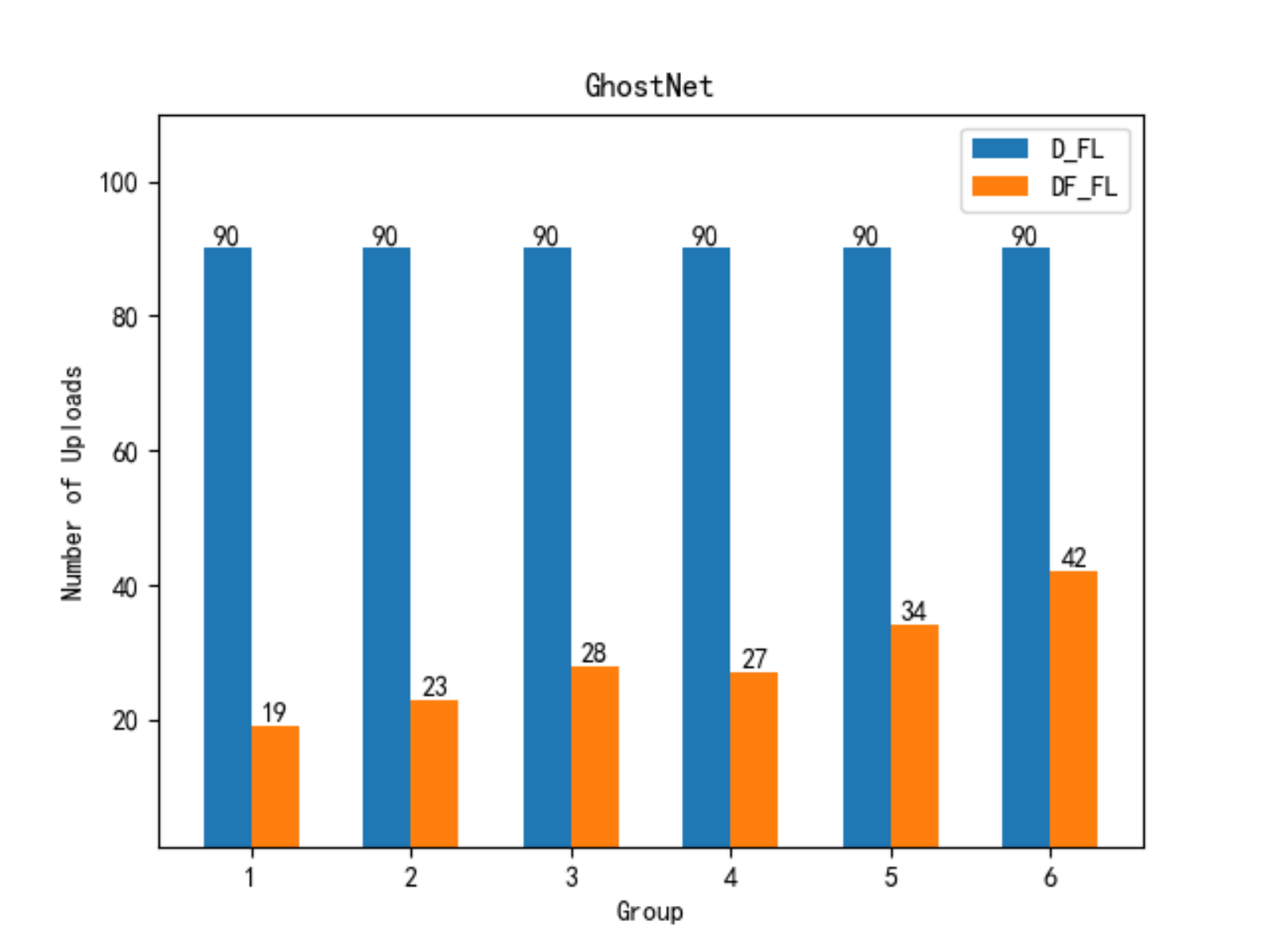}
    }
    \qquad
	\subfigure[ResNet50]{
	\centering
    \includegraphics[width=\columnwidth]{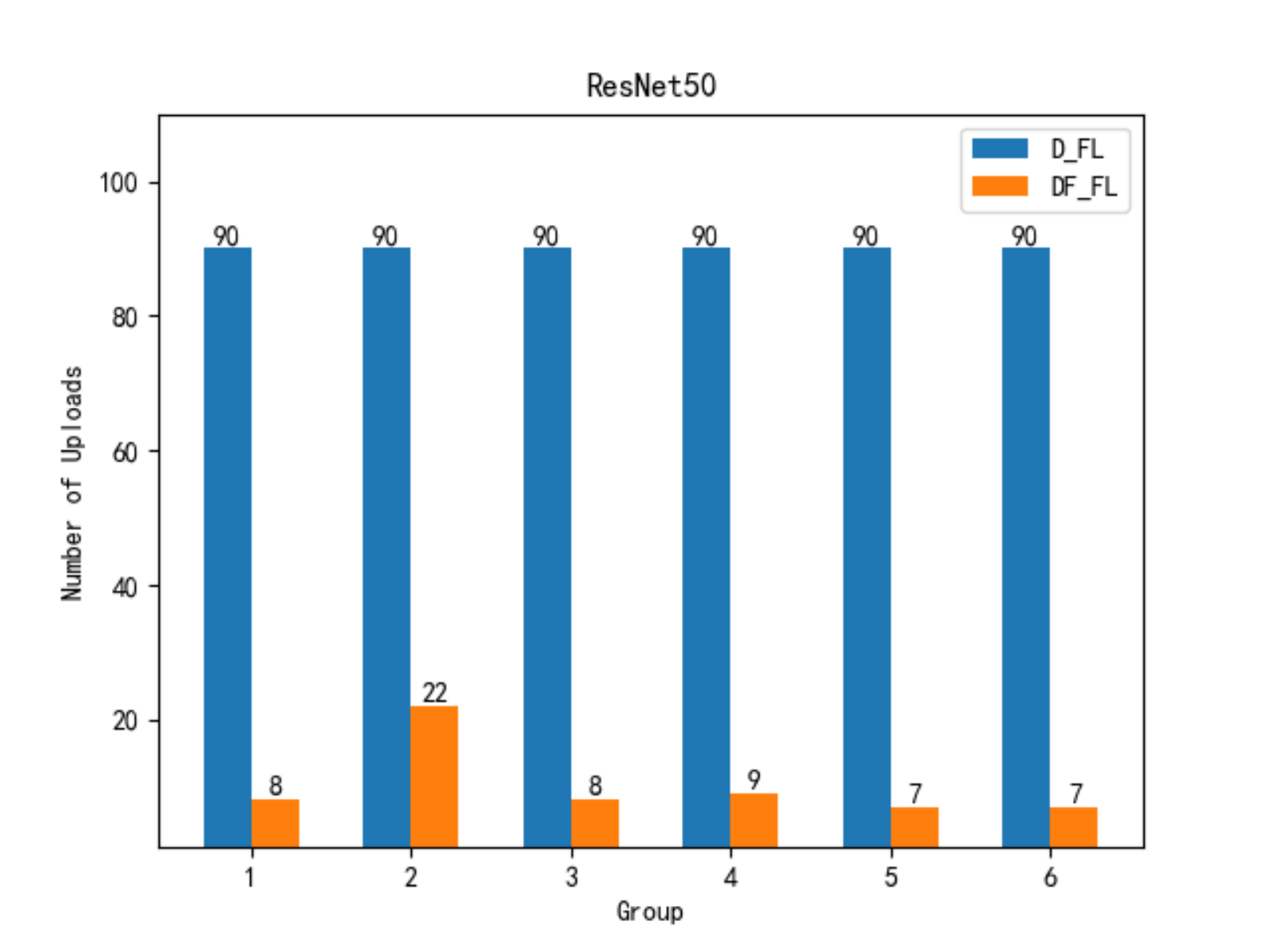}
    }
    \qquad
    \subfigure[ResNet101]{
    \centering
    \includegraphics[width=\columnwidth]{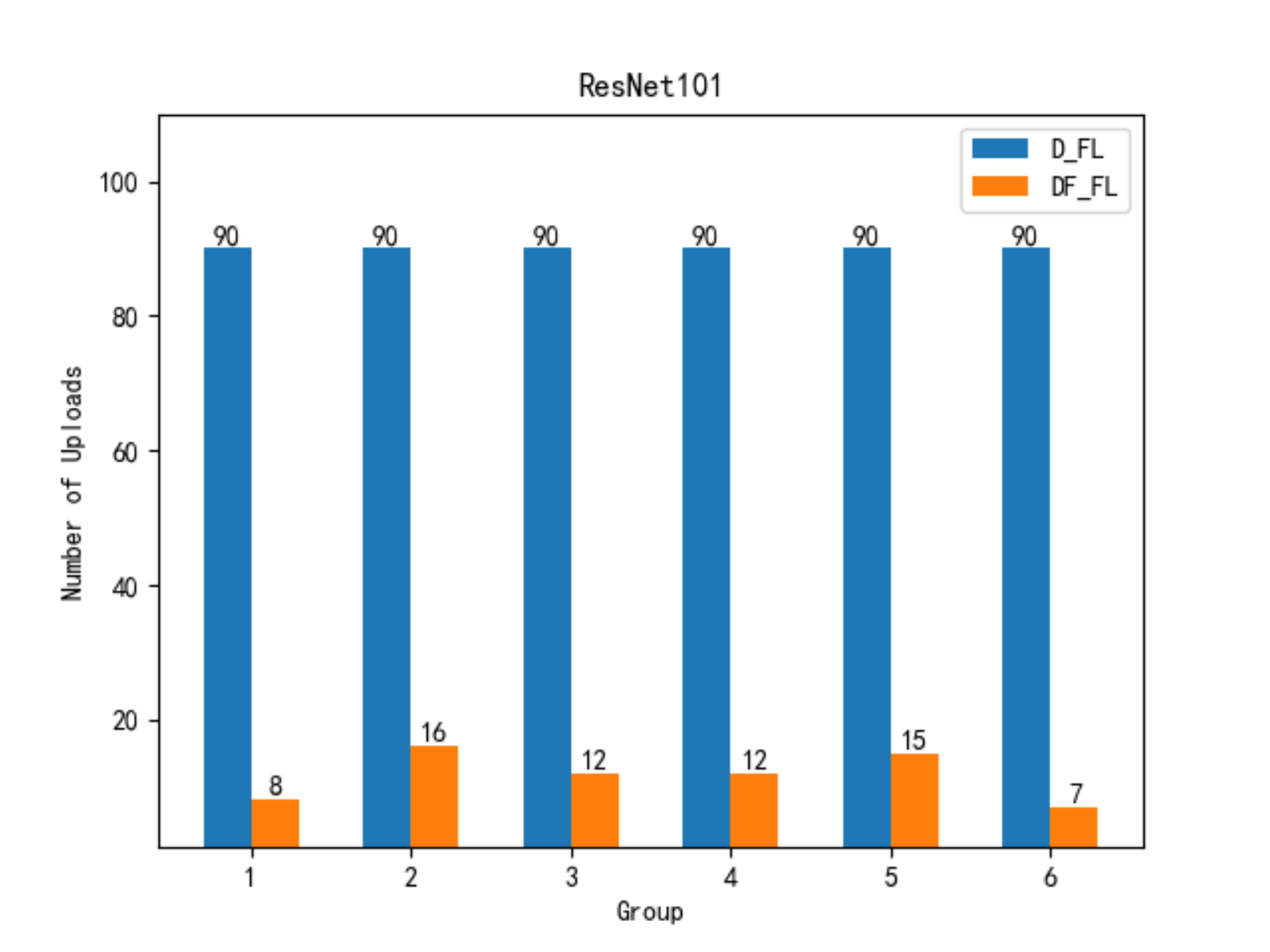}
    }
    \caption{Upload number.}
	\label{upnum}
\end{figure}

\begin{figure}[t]
	\centering
	\subfigure[GhostNet]{
	\centering
    \includegraphics[width=\columnwidth]{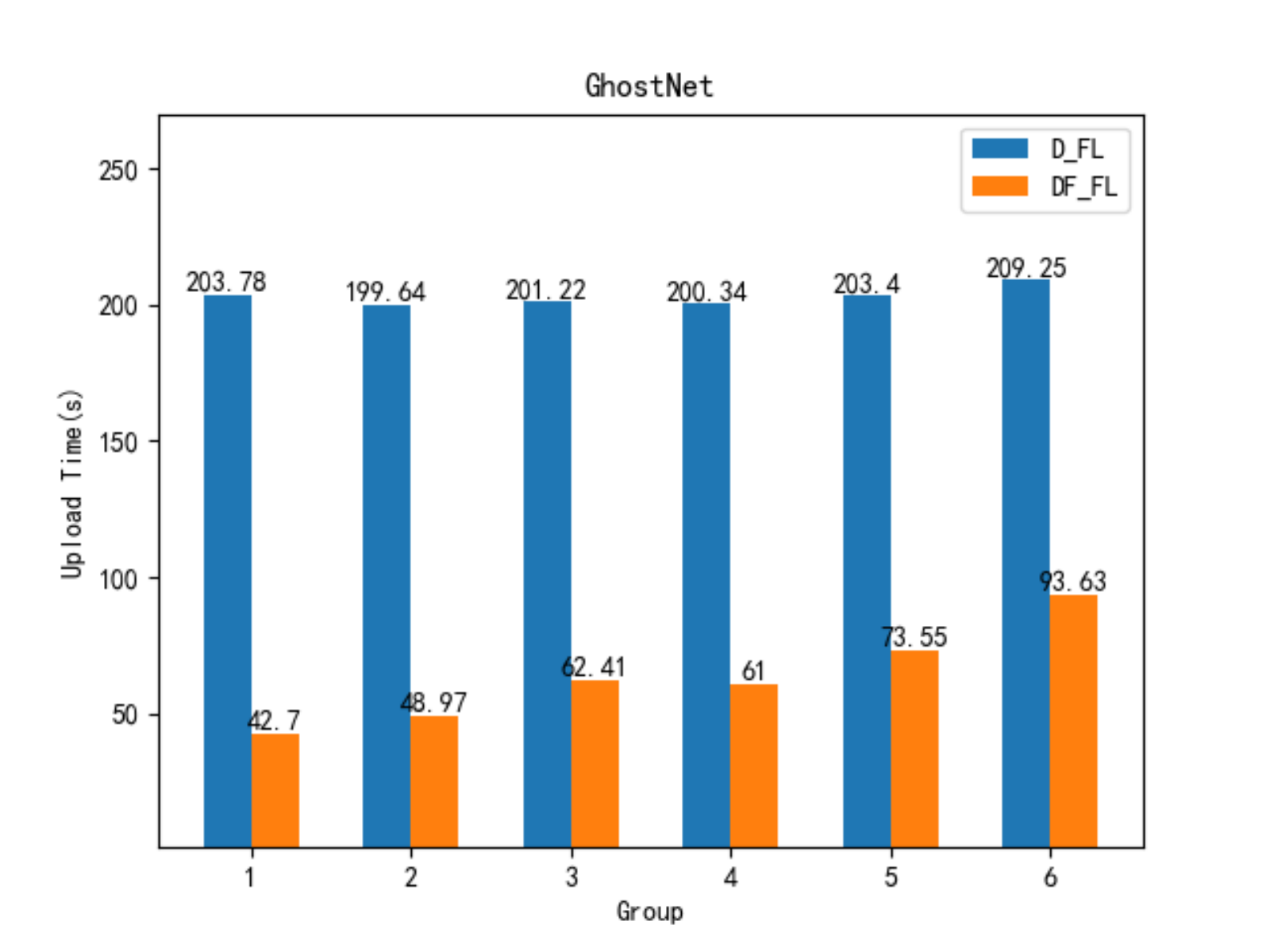}
    }
    \qquad
	\subfigure[ResNet50]{
	\centering
    \includegraphics[width=\columnwidth]{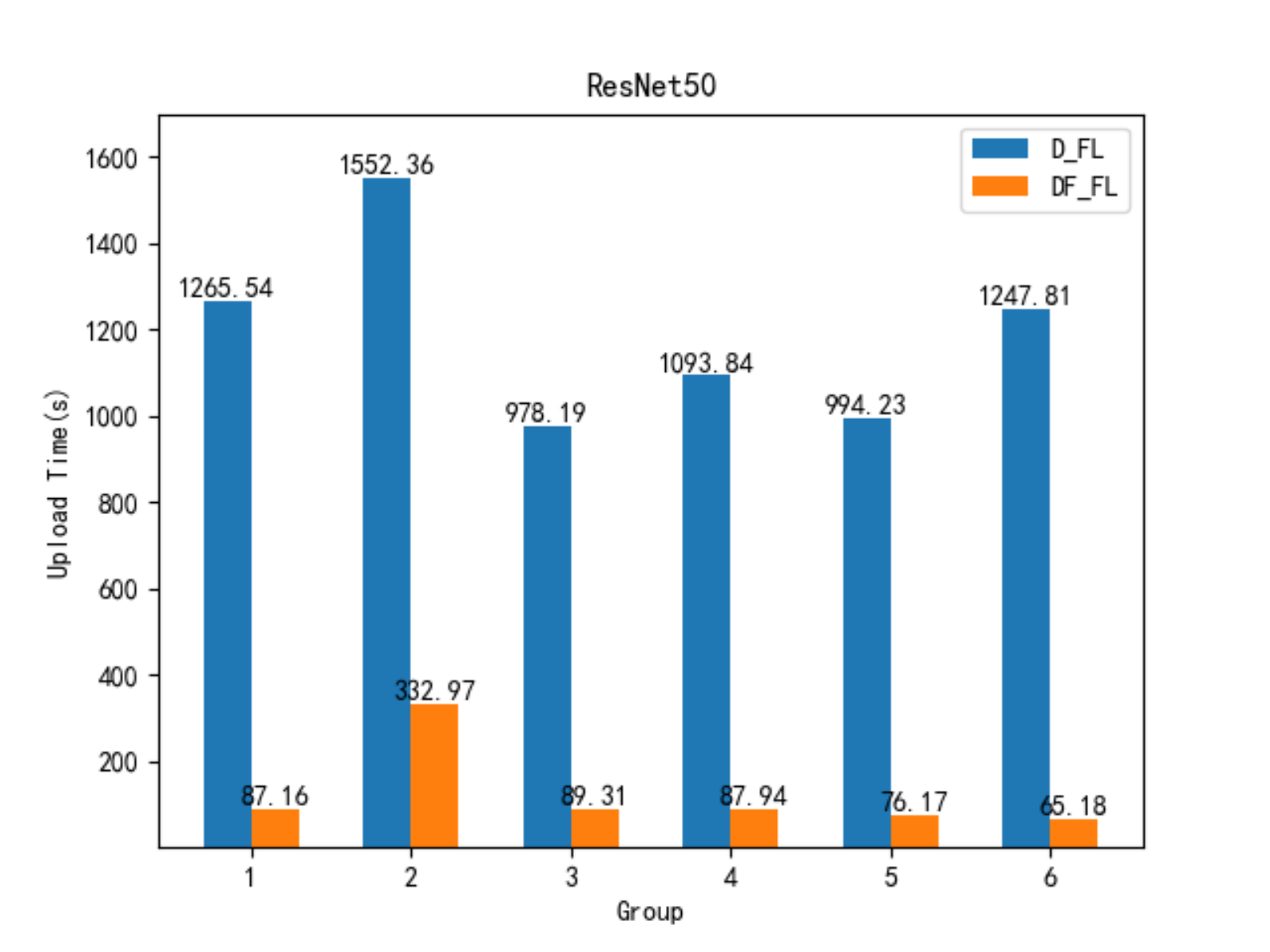}
    }
    \qquad
    \subfigure[ResNet101]{
    \centering
    \includegraphics[width=\columnwidth]{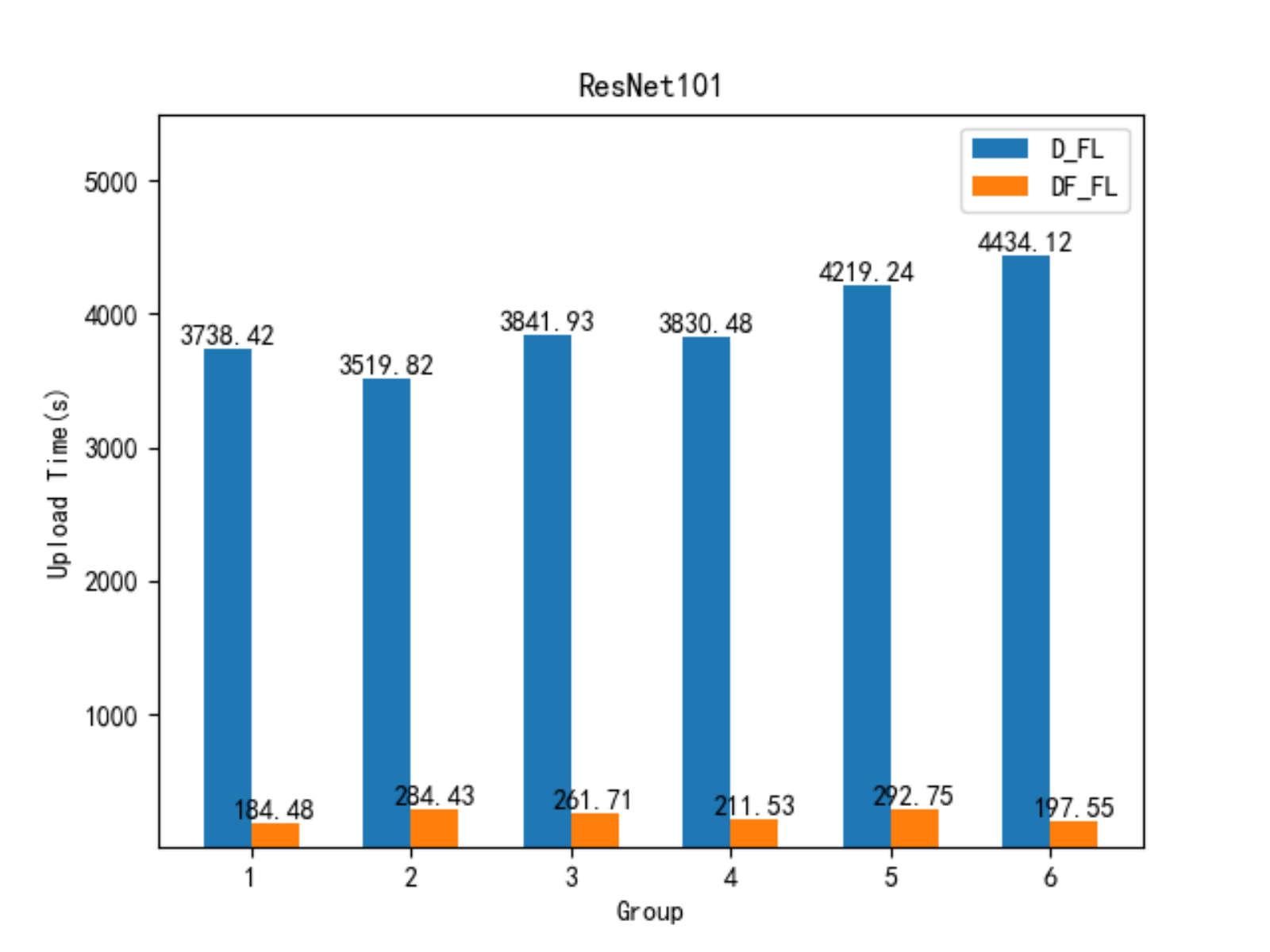}
    }
    \caption{Upload time.}
	\label{uptime}
\end{figure}

\subsection{Communication Efficiency}
To evaluate the effect of dynamic fusion on communication, we measure the upload number and upload time, which are shown in Fig.~\ref{upnum} and~\ref{uptime} respectively. Here the collected upload number and time are the total number of three clients, which in our case is 30 for each client and 90 in total.
In comparison with the default setting of federated learning (D\_FL) for GhostNet, the upload number of dynamic fusion is decreased by an average of 61, matching to a reduction of 110-160s of upload time (1/3 of the D\_FL time). For ResNet50, the upload number of dynamic fusion decreased by an average of 80, matching to a reduction of 900-1200s of upload time (1/10 of the D\_FL time). For ResNet101, the upload number of dynamic fusion decreased by an average of 78, matching to a reduction of 3200-4200s on upload time (1/16 of the D\_FL time).

Based on the results, we can conclude that dynamic fusion is capable to reduce the communication overhead through less model uploading. For models that have a simple structure and few parameters as GhostNet, the reduction is not significant (to only 1/3 of D\_FL). Nevertheless, dynamic fusion has more obvious effects in treating complicated models with more parameters (ResNet50 and ResNet101), which scale down to 1/10 and 1/16 of the D\_FL time.

\section{Related Work}
\label{background}
The concept of federated learning is first proposed by Google in 2016~\cite{DBLP:journals/corr/McMahanMRA16}, which initially focuses on cross-device learning. Google initially adopted federated learning to predict search suggestions, next words and emojis, and the learning of out-of-vocabulary words~\cite{DBLP:journals/corr/abs-1812-02903, chen2019federated, ramaswamy2019federated}. The scope of federated learning is then extended to cross-silo learning, e.g., for different organisations or data centers~\cite{kairouz2019advances, 2020SLR, Zhang2020}. For example, Sheller et al. \cite{10.1007/978-3-030-11723-8_9} build a segmentation model using brain tumor data from different medical institutions.

Although communication efficiency can be improved by only sending model updates instead of raw data, federated learning systems requires multiple rounds of communications during training to achieve model convergence. Many researchers work on the methods to reduce communication rounds~\cite{8917724, 8759317}. One way is through aggregation, e.g., selective aggregation~\cite{8994206}, aggregation scheduling~\cite{8851249}, asynchronous aggregation~\cite{xie2019asynchronous}, temporally weighted aggregation~\cite{8761315}, controlled averaging algorithms~\cite{karimireddy2019scaffold}, iterative round reduction~\cite{8917724}, and shuffled model aggregation~\cite{ghazi2019scalable}. Furthermore, model compression methods are utilised to reduce the communication cost that occurs during the model parameters and gradients exchange between clients and the central server~\cite{8885054}. Additionally, communication techniques are introduced to improve communication efficiency, e.g., over-the-air computation technique~\cite{8952884}, multi-channel random access communication mechanism~\cite{8935424}.  

Federated learning can
address statistical and system heterogeneity issues since models are trained locally~\cite{corinzia2019variational}. However, challenges still exist in dealing with non-IID data
Many researchers have worked on training data clustering~\cite{ sattler2019clustered}, multi-stage local training~\cite{jiang2019improving}, and multi-task learning~\cite{corinzia2019variational}. Also, some works ~\cite{8893114, 8733825} focus on incentive mechanism design to motivate clients to participate in the machine learning jobs. 

Federated learning has been recently adopted in CT or X-Ray image processing for COVID-19 positive case detection~\cite{liu2020experiments, kumar2020blockchainfederatedlearning}. However, the above studies do not consider the communication efficiency and model accuracy issues of federated learning. Our research work proposed a dynamic fusion-based approach to improve communication efficiency and model performance.

\section{Conclusion}
\label{conclusion}
This paper proposes a novel dynamic fusion-based federated learning approach to improve accuracy and communication efficiency while preserving data privacy for COVID-19 detection. 
The evaluation results show that the proposed approach is feasible and performs better than the default setting of federated learning in terms of model accuracy, fault tolerance, robustness, and communication efficiency. 



%





\ifCLASSOPTIONcaptionsoff
  \newpage
\fi

\end{document}